\documentclass[twocolumn,secnumarabic,amssymb, nobibnotes, aps, pre]{revtex4-1}
\usepackage{amsmath}
\usepackage{multirow}
\usepackage{array}
\usepackage[FIGTOPCAP]{subfigure}
\usepackage{graphicx}
\usepackage{tabularx}
\usepackage{epsfig}
\usepackage{epstopdf}
\usepackage{lipsum}

\begin{document}

\title{Pearling instability of membrane tubes driven by curved proteins and actin polymerization}
\author{U. Jeler\v ci\v c$^{1,2}$ and N. S. \surname{Gov}$^{2}$}
\email[Corresponding author: ]{nir.gov@weizmann.ac.il}
\affiliation{$^1$Jo\v zef Stefan Institute, Department of Theoretical Physics, Jamova 39, SI-1000 Ljubljana, Slovenia\\
$^2$Department of Chemical Physics, The Weizmann Institute of Science, P.O. Box 26, Rehovot, Israel 76100}

\begin{abstract}
Membrane deformation inside living cells is crucial for the proper shaping of various intracellular organelles and is necessary during the
fission/fusion processes that allow membrane recycling and transport (e.g. endocytosis). Proteins that induce membrane curvature play a
key role in such processes, mostly by adsorbing to the membrane and forming a scaffold that deforms the membrane according to the curvature
of the proteins. In this paper we explore the possibility of membrane tube destabilisation through a pearling mechanism enabled by the
combined effects of the adsorbed curved proteins and the actin polymerization they may recruit. The pearling instability can furthermore
serve as the initiation for fission of the tube into vesicles. We find that adsorbed proteins are more likely to stabilise the tubes, while
the actin polymerization can provide the additional constrictive force needed for the robust instability. We discuss the relevance of the
theoretical results to in-vivo and in-vitro experiments.
\end{abstract}

\maketitle

\emph{Introduction:} Intracellular membrane is continuously broken up into small vesicles needed for the transport and recycling of the cell
organelles \cite{watson2005er} and during endocytosis \cite{doherty2009mechanisms}. By adsorbing to the membrane and bending it, proteins that curve the membrane
play an important role during this processes \cite{sens2008biophysical,anitei2012bridging}. These proteins play a role during endocytosis (e.g. clathrin \cite{shevchuk2012alternative} and BAR-domain proteins \cite{meinecke2013cooperative}) and help during the later stages of fission and transport \cite{chavrier1999role} throughout the cell \cite{cai2014graf1}. In addition, the recruitment of actin is implicated as a key component allowing for the completion of vesicle fission \cite{miserey2010rab} and the endocytosis process \cite{anitei2012bridging}, especially in clathrin-independent pathway \cite{mayor2007pathways,romer2010actin,basquin2013signalling}. The mechanism enabling curved proteins to coordinate and drive the membrane vesiculation process is currently a subject of intense experimental and theoretical studies \cite{liu2009mechanochemistry,liu2010mechanochemical}. Recent in-vitro experiments have investigated the ability of curved proteins to induce the
formation of the membrane tubes, and it was found that the fission of the tubes into vesicles requires additional components such as actin
polymerization \cite{romer2007shiga,romer2010actin}, membrane demixing \cite{allain2004fission,roux2005role,sorre2009curvature} or dynamin-driven constriction \cite{morlot2013mechanics}. It is therefore still an open problem to understand the physical routes by
which curved proteins achieve vesicle formation, while it is rather understood how they form a scaffold that stabilizes membrane tubes
\cite{frost2008structural}. The coupling between curved membrane proteins and the recruitment of cytoskeletal components has been shown to
drive a large variety of membrane shapes \cite{itoh2005dynamin,gov2006dynamics,shlomovitz2008physical}.

The destabilization of a membrane tube due to adsorbed curved proteins is possible through the instability that leads to non-uniform
distribution of the curved proteins, as was explored in \cite{shlomovitz2009membrane}, and seen in cellular organelles such as golgi \cite{shemesh2003prefission}. In this mechanism large density undulations in the distribution of the adsorbed proteins lead
to regions that are constricted and rich in protein scaffold. While this process can play an important role, we wish to explore whether the
tube destabilisation is also feasible in the regime of uniform protein coverage, i.e. the adsorbed proteins form a rather dense scaffold. In this regime we investigate the pearling instability mechanism \cite{bar1994instability,yu2009pearling,boedec2014pearling}. Note that a previous study of pearling induced by curved membrane inclusions \cite{campelo2007model} treated grafted curved molecules, while we treat the case where the curved molecules are dynamically adsorbed from the surrounding medium. Finally, in parallel to our work, a recent study \cite{Zhang2014} simulated the details of the membrane shapes during tubulation and pearling for a model that closely resembles ours, and motivated by the fine details of the endocytosis process. Our study is aimed at the more general physics of this process.

We explore two possible routes by which curved concave proteins (of the F-BAR-like shape) can lead to the pearling of a membrane tube: (i) a pre-formed, bare membrane tube
is rapidly coated by the curved proteins; or (ii) a membrane tube is formed through the adsorption of the curved proteins, and is then squeezed
by the rapid polymerization of an actin coat. Note that for pearling to occur the factors that affect the membrane tube have to act faster than the rate at which the water can flow out of the tube. We find that only coating by curved proteins (mechanism (i)) is not robust, with most cases leading to \emph{increased} tube stability, while the addition of actin polymerization around the tube (ii) provides an extra squeezing force that can drive robust pearling instability.

\emph{Tubes with adsorbed curved proteins:} We begin with the equilibrium state of a membrane tube with adsorbed curved proteins. At
equilibrium, there is no net flux of proteins adsorbing and desorbing onto the lipid membrane. The equilibrium condition can be written as:
\begin{equation}
k_{ad}(1-\phi)\rho v=k_{de}\phi,
\label{eq:eq1}
\end{equation}
where $\rho=N_0/V$ is the density of the proteins in the surrounding solution (infinite reservoir), $\phi $ is the occupied area fraction,
$v$ is the volume of a single protein and $k_{ad}$ and $k_{de}$ denote the adsorption and desorption rates respectively, such that:
$0<\phi<1$ and $0<\rho v<1$. We additionally assume that the rates of adsorption and desorption depend on the protein and tube radius in the following manner \cite{peleg2011propagating}
\begin{equation}
k_{ad}=A_{f}\exp\left[{-\frac{\kappa_p a \beta}{2}\left(\frac{1}{R_t}-\frac{1}{R_p}\right)^2}\right]k_{de},
\label{eq:kad}
\end{equation}
where $\beta=1/k_BT$ and $\kappa_p$, $R_p$, $R_t$ nd $a$ represent the bending rigidity of the adsorbing protein molecule, the curvature
radii of the proteins and the lipid tube and the area of protein molecule, respectively (note that $R_t$ may in general not be constant).
The affinity, $A_{f}$, is a
numerical factor describing the preference of the proteins to adsorb to a specific surface. The adsorption will be more probable
($k_{ad}>k_{de}$) when both the protein and the tube radii match, whereas in the case of a large mismatch, the desorption represents the
dominating process. Finally, combining Eqs. (\ref{eq:eq1}) and ({\ref{eq:kad}}), we get the equilibrium number density of proteins on the
tube:
\begin{equation}
\phi=\left\{\frac{1}{A_{f}\rho v}\exp\left[\frac{\kappa_p a \beta}{2}\left(\frac{1}{R_t}-\frac{1}{R_p}\right)^2\right]+1\right\}^{-1}.
\label{eq:phi}
\end{equation}

\begin{figure}[!ht]
\centering
\includegraphics[width=\columnwidth]{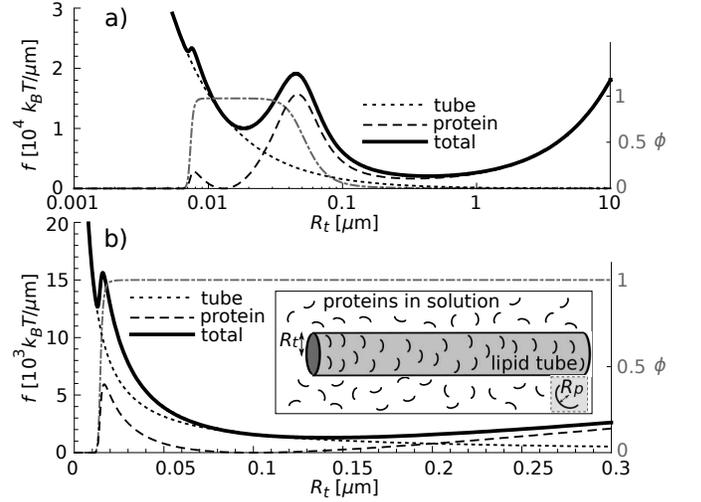}
\caption{Total free energy per unit length as a function of tube radius in the case of less (a,~$R_p~=~0.013\mu\text{m}$)  or more (b,
$R_p~=~0.1\mu\text{m}$) bent protein molecules and at vanishing $\sigma$. Dotted and dashed lines represent the tube and protein
contribution to the total free energy and the dash-dotted gray line shows the dependence of occupied area fraction on the tube size
($\kappa_p\beta=\kappa_t\beta=50$, $\rho=5\cdot10^5/\mu\text{m}^3$, $a=10^{-4}\mu\text{m}^2$, $v=10^{-6}\mu\text{m}^3$, $A_f=10^4$ and
$p=0$).}
\label{fig:fUni1}
\end{figure}

Let us illustrate this result in the case of a uniform lipid tube of fixed radius $R_t$. If the curvature of the protein molecules is
large, $R_p\ll\sqrt{a\beta\kappa_p}$  (Fig. \ref{fig:fUni1}a), the occupied area fraction of adhered proteins increases steeply when $R_t$
approaches $R_p$, and peaks at $R_t\approx R_p$. The occupation fraction then decreases rapidly as the tube radius is further increased. In
the limit of $R_t\rightarrow\infty$ the occupation fraction reaches a finite value, which approaches zero for $R_p\rightarrow0$
(Eq.\ref{eq:phi}). The curve $\phi(R_t)$ is asymmetric around the value $R_t=R_p$, which becomes more pronounced for less bent proteins
($R_p\gg\sqrt{a\beta\kappa_p}$, Fig. \ref{fig:fUni1}b). In this case $\phi$ again increases sharply as $R_t$ approaches $R_p$ from below,
but the occupied area fraction saturates much before $R_t=R_p$, and remains roughly constant as $R_t\rightarrow\infty$, approaching
$\phi_{\infty}=\left[(A_{f}\rho v)^{-1}+1\right]^{-1}$ for $R_p\rightarrow\infty$ ($\phi_{\infty}\rightarrow1$, for the strong affinity case which is of interest here, $A_{f}\rho v\gg1$). In this limit of $R_p\rightarrow\infty$ the weakly curved proteins do not discriminate greatly between differently curved lipid tubes.

A uniform lipid tube that is coated uniformly by the adsorbed proteins has a different equilibrium radius compared to the bare membrane. The
new tube radius is found by minimizing the total free energy of the system (per unit length), which is now comprised of the bending energy
of the tube, the bending energy of the proteins and the terms associated with the volume/area or pressure/surface tension conservation \cite{sorre2009curvature}
\begin{equation}
f=\left[\frac{\kappa}{R_t^2}+\kappa_p\left(\frac{1}{R_t}-\frac{1}{R_p}\right)^2\phi+2\sigma-p R_t\right]\pi R_t,
\label{eq:f}
\end{equation}
and where $\phi$ is given by the Eq. (\ref{eq:phi}), $\kappa$ is the bending modulus of the membrane, $\sigma$ is the effective membrane
tension and $p$ is the osmotic pressure.

\begin{figure}[!ht]
\centering
\includegraphics[width=\columnwidth]{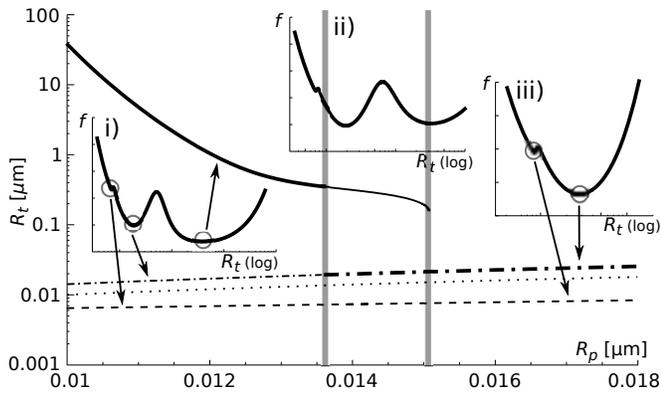}
\caption{The local and the global minima of the total free energy per unit length as a function of curvature of the protein molecules (the
dotted line represents the $R_p=R_t$ line). The insets (where we schematically plot the total free energy as a function of the tube radius)
show three distinct cases: i) at low enough $R_p$ we have three minima --- one at $R_t\ll R_p$ (dashed), one at $R_t\sim R_p$ (dash-dotted)
and a global one at $R_t\gg R_p$ (solid), ii) at certain $R_p$ the last two minima become degenerate and iii) at large $R_p$ there exists
only a local minimum at $R_t\ll R_p$ and a global one at $R_t\sim R_p$. The global minimum in the whole range of $R_p$ is marked in bold.
Note that the affinity has to increased enough in order to realise all the minima. The system parameters here are the same as in the Fig.
\ref{fig:fUni1}.}
\label{fig:minLogA}
\end{figure}

Fig. \ref{fig:fUni1} shows the free energy per unit length in the case of $\sigma=p=0$ for two limiting cases. i) For weakly curved
proteins, the free energy has two minima and the global one is found at: $R_t\simeq R_p(\kappa_p\phi+\kappa)/\kappa_p\phi$. This is easily
seen from Eq. (\ref{eq:f}) in the limit of large $R_p$, where $\phi\simeq1$ (Figs. \ref{fig:fUni1}b, \ref{fig:minLogA}). b) For highly
curved proteins, there are now two local minima and the global minimum can shift to very large values $R_t\gg R_p$. In this limit
(Figs.\ref{fig:fUni1}a, \ref{fig:minLogA}) the adsorbed proteins act mainly as an additional membrane tension, and $R_t\simeq
R_p\sqrt{(\kappa_p\phi+\kappa)/\kappa_p\phi}$, where $\phi\ll1$ (Eq.\ref{eq:f}). The global minimum at large tube radius is eventually
influenced by the finite membrane tension. Note that the local minima in both cases exist only at sufficiently high $A_{f}\rho v$.

These results can be further generalized by adding a non-zero (positive) surface tension, which non-uniformly raises the free energy curve
thus making it possible for the minima to change their characters (from local to global and vice versa) and causes not only double but
possibly also triple energy degeneration. By varying the surface tension any given tube radius can be achieved.

\emph{Pearling of pre-formed membrane tubes:} Let us now a cylindrical pure membrane tube (equilibrated at $p=0$ and $\sigma>0$) with an
equilibrium radius $R_t=\sqrt{\kappa/2\sigma}$. Curved proteins are then introduced into the surrounding medium and the tube is quickly
coated. If the coating process is slow, so that the solution inside the tube has time to flow and adjust the tube volume, the adsorbed
proteins drive the tube to change its radius uniformly, approaching the equilibrium tube radius calculated in the previous section (Fig.
\ref{fig:minLogA}). If the coating process is very fast, the tube has a fixed volume and can only change its surface area (shape). In this
limit of fixed volume the tube can only lower its free energy by modifying shape and a possible alternative morphology is a sinusoidally
modulated tube (pearled shape), with radius parametrized as \cite{bar1994instability}: $R(z)=R_m+\epsilon\sin(q z)$, where $R_m$, $\epsilon$, $q$ and $z$
denote the mean radius, amplitude of the perturbation, wave vector of the perturbation and the coordinate along the cylindrical axis,
respectively.

\begin{figure}[!ht]
\centering
\includegraphics[width=\columnwidth]{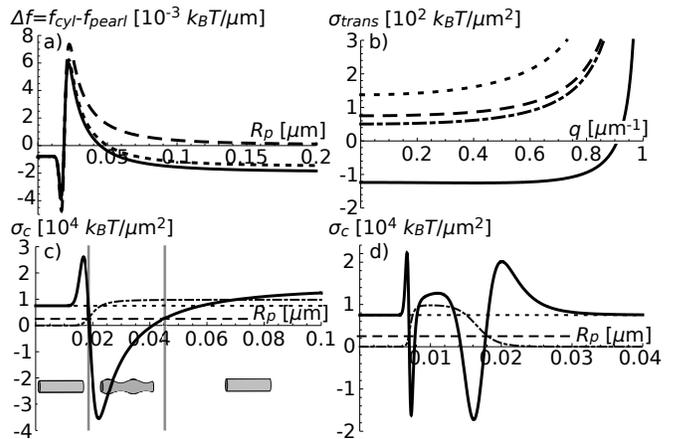}
\caption{(a) The difference between the free energy per unit length of the initial cylinder (equilibrated at $\sigma_0=2500\mu
k_BT/\text{m}^2$) and the pearled shape with the same volume and length in the limit of $q,\epsilon\rightarrow~0$ as a function of
protein curvature. The curves are plotted at $\sigma_a=0$, $2500$, $12500$ $k_BT/\mu\text{m}^2$ (solid, dotted and dashed lines,
respectively). (b) The surface tension required for the transition from the cylinder to the pearled state at several $R_p$ ($0$, $0.3$, $0.5$,
$1.5$ $\mu\text{m}$; dash-dotted, solid, dashed and dotted lines, respectively) as a function of the wave vector at $\sigma_0=25\mu
k_BT/\text{m}^2$. (c) Critical surface tension (solid line), needed for the phase transition in the limit of $q,\epsilon\rightarrow~0$ and at
$R_t=0.1\mu\text{m}$ and (d) $R_t=0.01\mu\text{m}$ (in both plots the dotted, dashed and dash-dotted lines represent the limit without
proteins, the initial surface tension needed to form the tube and the area occupation fraction, respectively). The system
parameters here are the same as in the Fig. \ref{fig:fUni1}, except for $A_f=100$.}
\label{fig:sigmaA}
\end{figure}

Since the volume of the pearled shape is kept fixed (equal to the volume of the initial tube, $R_t$), the following relation holds:
$R_m=\sqrt{R_t^2-\epsilon^2/2}$. The free energy can now be calculated by integrating the Eq. (\ref{eq:f}) over a unit length and dividing
it by the same unit, taking into account that the tube radius is a function of $z$ ($R(z)$). We limit ourselves to the case where
$R_m\gg\epsilon$, but nevertheless the normalised free energy is too complicated to be written here. Since we find that the free energy
does not depend greatly on the wave vector (except at very large $q$, Fig. \ref{fig:sigmaA}b), we additionally concentrate on the limit
$q\rightarrow0$.

We compare the normalized free energies of the coated pearled tube and the coated uniform tube (at the initial radius $R_t$), at $q=0$. We
find that the equilibrium shape depends strongly on the protein curvature (Fig. \ref{fig:sigmaA}a, solid line). The regime where the
cylindrical shape has a higher energy than the pearled shape ($\Delta f>0$), is the regime of spontaneous pearling. In Fig.
\ref{fig:sigmaA}b we plot the critical tension needed to achieve pearling $\sigma_{trans}$ (e.g. for the energy difference between pearled
and uniform tube to vanish), as a function of the wavevector $q$, and for different values of the protein curvature radius. We find that
weakly curved proteins (large $R_p$) act to stabilize the tube, raising the value of $\sigma_{trans}$ (lower energy for the cylindrical
shape in Fig. \ref{fig:sigmaA}a). More strongly curved proteins can lower $\sigma_{trans}$, and even drive it to negative values, thereby
ensuring spontaneous pearling (Fig. \ref{fig:sigmaA}b).

We now evaluate the critical surface tension at $q=0$ (Fig.\ref{fig:sigmaA}c, d), $\sigma_c=\sigma_{trans}(q=0)$ and then compare it to the
equilibrium surface tension $\sigma_0$ needed to create and stabilize the initial lipid tube (prior to the introduction to the proteins). We
find that, depending on the mismatch between the initial tube radius and the protein curvature, $\sigma_c$ can be either bigger or smaller
than $\sigma_0$. If we first focus on the Fig. \ref{fig:sigmaA}c ($R_t=0.1\mu\text{m}$), we see that with increasing $R_p$ the critical
surface tension first rapidly increases ($\sigma_{c}>\sigma_0$), then drops sharply to low (even negative) values ($\sigma_{c}<\sigma_0$),
and then monotonously increases as the protein curvature decreases. The initial rise in $\sigma_c$ at low $R_p$ is due to a low coverage by
the proteins, which have a large mismatch between $R_p$ and $R_t$ (Fig.\ref{fig:fUni1}b). The region where the critical tension is lowered
is a consequence of the large increase in the protein coating of the tube and a relatively large mismatch between $R_p$ and $R_t$ favouring
the shape change. Finally, for larger $R_p$ the occupation fraction approaches a constant value and the critical surface tension gradually
increases. In this limit ($R_p\rightarrow\infty$) the protein coat simply acts to stiffen the tube by increasing the effective bending
modulus [Eqs. (\ref{eq:phi}), (\ref{eq:f})], and thereby stabilizing the tube against pearling.

At very low tube radii the critical surface tension exhibits additional spikes (Fig. \ref{fig:sigmaA}d), due to the additional sharp drop in
the occupation fraction (Fig. \ref{fig:fUni1}a), when $R_p\rightarrow\infty$. The bulk protein density or affinity do not have a strong
qualitative effect and on our results, except in the limit of very small density of the proteins in the solution or very low affinity.

Note that given enough time the condition of conserved volume does not apply anymore and the system will relax towards the lowest energy
state, which is (at any given $R_p$) a cylindrical tube with a readjusted radius (Fig. \ref{fig:minLogA}). The timescale for the volume
readjustment by water displacement flow depends on the overall change in volume, being faster for shorter tubes and for smaller changes in
tube radius. As we show in Fig. \ref{fig:minLogA}, the equilibrium radius of the protein-coated tube is (in the majority of the domain)
close to $R_p$, which means that the biggest jumps in radius compared to the initial radius happen at low $R_p$, where the tube has regimes
where it can pearl spontaneously. Since the tube can not change its radius fast when the jump in radius is large, the cylindrical membrane
will most likely satisfy the conditions for pearling instability upon being coated with the proteins. However, we find that the conditions
for spontaneous pearling are limited to very specific "instability windows" of $R_p$ (Fig. \ref{fig:sigmaA}c, d), which makes this mechanism non-robust, as
most values of $R_p$ end up stabilizing the membrane tube.

\emph{Pearling of protein-coated membrane tubes due to actin polymerization:} We now address the effect of actin polymerization on the
pearling instability. For simplicity we consider that the same curved proteins treated above can recruit the nucleation of actin
polymerization, even though in reality these functions are linked by a variety of proteins that act together. The actin filaments that
polymerise against the membrane tube can create an inwards squeeze on the tube, thereby acting as an additional effective membrane tension
that can drive the pearling instability. When actin polymerizes against the membrane, the treadmilling motion of the filament against the
surrounding viscous network produces an effective friction force that converts the polymerization into a pushing force acting normal to the
membrane (on average) \cite{orly2014physical}. Alternatively, an expanding coat of cross-linked actin gel build up internal stress, proportional to the curvature of the membrane surface \cite{noireaux2000growing}: assuming that the actin gel is relaxed when forming on the membrane tube, it builds up the following internal tension when pushed to the edge of the actin coat: $\sigma\propto k (l_0w/R_t)$, where $w$ is the actin coat thickness (determined by the polymerization and severing/depolymerization rates), $l_0$ is the separation between actin filaments at the membrane
surface and $k$ is the effective elastic spring constant for the cross-linked actin gel. We therefore expect that the actin-induced tension
will increase with the tube curvature, until it reaches a maximal value given by the stall force of actin polymerization, of the order of
$1\text{nN}$ per filament \cite{footer2007direct}. Note that if myosin-II is also recruited to the actin coat, an additional squeezing force can arise from myosin-driven contractility \cite{barfod2011myosin}.

We therefore include the protein-dependent surface tension contribution into the free energy by adding a term $\sigma_a\phi$ into the
Eq. (\ref{eq:f}). This additional tension (together with the existing tube tension) leads to a more robust pearling due to the formation of
an protein-actin coat on a pre-formed tube (Fig. \ref{fig:sigmaA}a). Note that above a critical value of $\sigma_a$ the pearling
regime extends to all values of $R_p$ (above some minimal threshold).

\begin{figure}[!ht]
\centering
\includegraphics[width=\columnwidth]{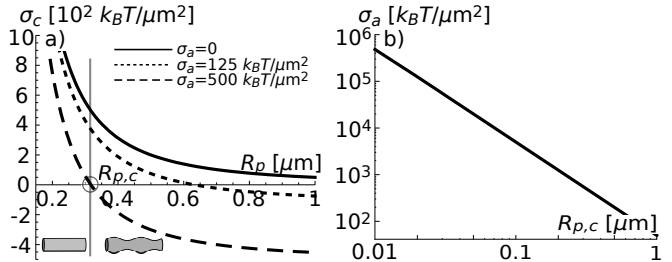}
\caption{(a) Critical surface tension necessary for pearling (for $q, \epsilon\rightarrow0$) at different $\sigma_a$ in
the case where the tube is formed by the curved proteins coat (at vanishing external surface tension), but the polymerization is triggered later. For the largest value of $\sigma_a$ we denote the vanishing point by $R_{p,c}$, whereby larger protein radii lead to spontaneous pearling. (b) The polymerization surface tension (at vanishing external surface tension) required for the spontaneous
pearling to occur, as a function of the critical protein curvature $R_{p,c}$. The system parameters here are the same as in the Fig. \ref{fig:fUni1}.}
\label{fig:sigma1}
\end{figure}

Let us now consider a slightly different scenario, whereby the protein coating is initiating the formation of the tube in the first place,
i.e. at a vanishing initial membrane tension. The tube radius of the initial cylinder depends now very strongly on the curvature of the
proteins, having usually a radius of the same order, $R_t\sim R_p$ (Fig. \ref{fig:minLogA}). We calculate the critical membrane tension
needed to induce the pearling instability $\sigma_c$, and we find that for each value of $\sigma_a$ there is a value of $R_p$ above which
the tube is spontaneously pearled due to the actin-induced tension ($\sigma_c<0$ in Fig. \ref{fig:sigma1}a). In Fig. \ref{fig:sigma1}b we
plot $\sigma_a$ versus the critical value $R_{p,c}$ above which the tube is destabilized, and we find that $\sigma_a\propto 1/R_{p,c}^{2}$: as expected, thinner tubes are harder to destabilize. The critical values of the actin-induced tension shown in
Fig. \ref{fig:sigma1}b are calculated using realistic values for the membrane and protein bending moduli. From the stall force of actin we
expect a maximal force of $\sim10^4\text{nN}/\mu\text{m}^2$ (using a maximal density of one actin filament per $100\text{nm}^2$), which is
much larger than the value estimated in Fig. \ref{fig:sigma1}b for the smallest protein radius: $\sigma_a/R_p\sim10^5$
[$k_BT/\mu\text{m}^2$]/$0.01$ [$\mu\text{m}$]$\sim50\text{nN}/\mu\text{m}^2$. We therefore conclude that actin polymerization in a dense gel
can supply enough squeezing force to drive the pearling of even the thinnest tubes.

\emph{Conclusion:} We have shown here that curved membrane proteins can both drive the formation of membrane tubes and destabilize them to
undergo pearling transition. This transition can be the first stage towards membrane fission and vesicle formation. However, we find that
the curved membrane proteins are more likely to stabilise the tubes and that for robust pearling an additional mechanism is necessary. This
can be in the form of recruitment of actin polymerization by the curved proteins, and subsequential growth of the actin gel which induces a
squeezing force that drives the pearling instability. Once the regions of constricted tube are thin enough, dynamin can be recruited \cite{roux2010membrane} to complete the vesicle scission. These results may help shed light on experimental observations both in-vitro and
in-vivo, where curved membrane proteins have been found to play key roles during membrane deformation, fission and transport. Previously the pearling instability was shown to occur in cellular membranes only in pathological cases \cite{bar1999pearling,marzesco2009release}, and we propose here that it may provide a mechanism used by cells for normal membrane remodeling.


\begin{acknowledgments}
UJ wishes to thank MFA Israel for their generous scholarship, NSG wishes to thank the Pasteur-Weizmann Council for their generous support,
and to Nathalie Sauvonnet for useful discussions. This research is made possible in part by the historic generosity of
the Harold Perlman Family.
\end{acknowledgments}

\bibliographystyle{apsrev4-1}
\bibliography{PearlingBib}

\end{document}